\documentclass[epj,nopacs]{svjour}

\usepackage{latexsym}
\usepackage{graphics}
\newcommand{\etal}{\textit{et al.}}
\journalname{}
\begin{document}
\title{Efficient production of an $^{87}$Rb $F=2$, $m_F=2$ Bose-Einstein condensate in a hybrid trap}
\titlerunning{Efficient production of an $^{87}$Rb $F=2$, $m_F=2$ BEC in a hybrid trap}
\author{Hari Prasad Mishra \and Adonis Silva Flores \and Wim Vassen \and Steven Knoop\thanks{e-mail: \texttt{s.knoop@vu.nl}}}
\institute{LaserLaB, Department of Physics and Astronomy, VU University, De Boelelaan 1081, 1081 HV Amsterdam, The Netherlands}
\date{\today}

\abstract{
We have realized Bose-Einstein condensation (BEC) of $^{87}$Rb in the $F=2$, $m_F=2$ hyperfine substate in a hybrid trap, consisting of a quadrupole magnetic field and a single optical dipole beam. The symmetry axis of the quadrupole magnetic trap coincides with the optical beam axis, which gives stronger axial confinement than previous hybrid traps. After loading $2\times 10^6$ atoms at 14~$\mu$K from a quadrupole magnetic trap into the hybrid trap, we perform efficient forced evaporation and reach the onset of BEC at a temperature of 0.5~$\mu$K and with $4\times10^5$ atoms. We also obtain thermal clouds of $1\times10^6$ atoms below 1~$\mu$K in a pure single beam optical dipole trap, by ramping down the magnetic field gradient after evaporative cooling in the hybrid trap.
}

\maketitle
\section{Introduction}\label{intro}

The experimental realization of Bose-Einstein condensates (BEC) of dilute atomic gases \cite{anderson1995oob,davis1995bec} is most often based on laser cooling and subsequent evaporative cooling in magnetic or optical dipole traps, or both, in either a sequential or combined way. A simple approach to achieve BEC is a hybrid trap that consists of a single beam optical dipole trap (ODT) and a quadrupole magnetic trap (QMT) \cite{lin2009rpo}. This hybrid trap combines the most simple magnetic trap and optical dipole trap in a way that one benefits from their individual strengths, i.\,e.\, a large trap volume to capture the laser-cooled cloud of atoms, tight confinement and efficient evaporation, while minimizing their weaknesses, i.\,e.\, Majorana spin-flip losses in a QMT and a small trap volume of an ODT. After evaporative cooling one can simply transfer the ultracold sample (or BEC) to a pure ODT by switching off the QMT completely. An experimental advantage over all-optical cooling methods (see e.\,g.\, Refs.~\cite{clement2009aor,olson2013ote,jiang2013sae}) is the much lower ODT power needed for the hybrid trap.

The hybrid trap has been successfully applied in several experiments, for $^{87}$Rb \cite{lin2009rpo,cho2011ahp,jenkin2011bec,gotlibovych2012acs,xiong2013narbbec,xiong2013poa,kuhn2014abc,lundblad2014ool}, $^{85}$Rb \cite{marchant2012bec}, $^{133}$Cs \cite{jenkin2011bec} and $^{23}$Na \cite{xiong2013narbbec}. In all these previous hybrid traps the symmetry axis of the QMT is placed vertically, while the ODT is in the horizontal plane, and forced evaporative cooling in the QMT is done by RF radiation. For $^{87}$Rb all experiments are done for the $F=1$, $m_F=-1$ hyperfine substate. 

Here we report on the realization of BEC of $^{87}$Rb in the $F=2$, $m_F=2$ hyperfine substate, in a hybrid trap in which the axial axes of both the QMT and ODT cross under a small angle in the horizontal plane, and forced evaporative cooling in the QMT is done by microwave (MW) radiation. Our choice for the $F=2$, $m_F=2$ state is primarily motivated by the suppression of inelastic collisions for an ultracold mixture of Rb and metastable triplet helium \cite{byron2010sop,knoop2014umo}, similar to the case of other mixtures with Rb \cite{modugno2001bec,roati2007bec,silber2005qdm,marzok2007uto,taglieber2008qdt}. It also provides a stronger confinement than the $F=1$, $m_F=-1$ hyperfine substate, as in the QMT the peak density scales with the magnetic moment to the third power. Furthermore, our orientation of the QMT with respect to the ODT allows for a four times stronger axial confinement, providing an additional enhancement of the peak density. The use of MW radiation for evaporative cooling is also motivated by the application of an atomic mixture, as MW-induced forced evaporative cooling is species-selective. Several groups have reported on the unwanted appearance of atoms in the $F=2$, $m_F=1$ state during the MW-induced forced evaporation in harmonic magnetic traps \cite{silber2005qdm,marzok2007uto,taglieber2008qdt,haas2007ssm,xiong2010ecr,wang2011eio}. We have performed Stern-Gerlach imaging to investigate the spin purity of our sample.

This paper is organized as follows. In Sect.~\ref{HT} we introduce the hybrid trap, and derive a simple analytic formula for the density profile. In Sect.~\ref{expsetup} we describe our experiment and in Sect.~\ref{results} we give our experimental results, discussing the alignment and loading of the hybrid trap, evaporative cooling towards BEC, the spin purity, and transfer to a pure ODT. Finally, we conclude and give an outlook in Sect.~\ref{conclusions}. 

\section{Hybrid trap}\label{HT}

\begin{figure}
\center
\resizebox{0.45\textwidth}{!}{%
  \includegraphics{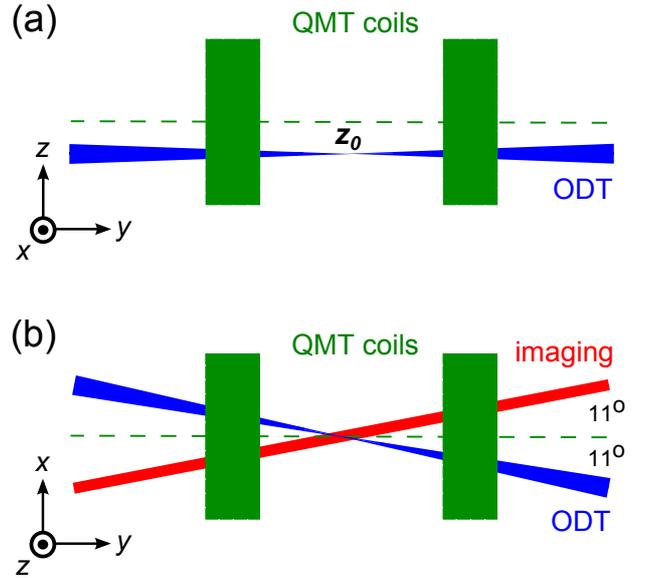}
}
\caption{(Color online) Schematics of our hybrid trap configuration (QMT and ODT), (a) showing the offset $z_0$ in the $y-z$ plane and (b) showing the angles between the QMT axis, ODT beam and absorption imaging beam in the $x-y$ (horizontal) plane.}
\label{setup}       
\center
\end{figure}

In the hybrid trap, as realized by Lin \etal~\cite{lin2009rpo}, a single beam ODT is aligned below the QMT center (see Fig.~\ref{setup}(a)), such that the trap minimum of the combined magnetic and optical trap is at a finite magnetic field, and atoms do not suffer Majorana spin-flip losses. After RF- or MW-induced forced evaporative cooling in the QMT the magnetic field gradient of the QMT is ramped down to the levitation gradient $B'_{\rm lev}\equiv m g / \mu$, where the vertical gradient compensates gravity. Here $m$ is the mass, $g$ is the gravitational acceleration and $\mu=g_F m_F \mu_B$ is the magnetic moment of the atom, where $g_F$ is the Land\'{e} factor of the hyperfine state $F$, $m_F$ is the quantum number of the Zeeman state, and $\mu_B$ is the Bohr magneton. Lowering the power in the ODT beam allows further (one-dimensional) evaporative cooling in the hybrid trap, in which the hot atoms can escape mainly downwards. An extensive analysis of the hybrid trap, in particular the transfer from the QMT to the hybrid trap, is given in Sect.~II of Ref.\,\cite{lin2009rpo}. Here we summarize the main ingredients, with the aim to provide a simple analytic expression of the density profile in the hybrid trap.  

The trapping potential of the hybrid trap is given by:
\begin{eqnarray}
U(x,y,z)&=&\mu B' \sqrt{x^2+4y^2+(z-z_0)^2} \\
&&-\frac{2PC}{\pi w(y)^2}\exp\left[-2\frac{x^2+z^2}{w(y)^2}\right]+m g z \nonumber,
\end{eqnarray}
where the first term is the QMT potential, the second term the ODT potential, and the third term the gravitational potential. In our case, the symmetry (strong) axis of the QMT and the ODT beam are along the $y$-axis, the $z$-axis is the vertical direction (see Fig.~\ref{setup}(a)). Here $B'$ is the magnetic field gradient along the weak axis of the QMT, $z_0$ is the vertical displacement of the QMT with respect to the ODT (such that a positive $z_0$ means that the ODT is placed below the QMT center), $P$ is the power of the ODT beam, $C=\alpha_{\rm pol}/2\epsilon_0 c$ is a constant proportional to the polarizability $\alpha_{\rm pol}$, depending on the atomic species and used wavelength $\lambda$, $w(y)=w_{0}\sqrt{1+y^2/y_{R}^2}$, where $w_{0}$ and $y_{R}=\pi w_{0}^2/\lambda$ are the beam waist ($1/e^2$ radius) and the Rayleigh length, respectively.

For temperatures much smaller than the trap depth, the trapping potential can be approximated by:
\begin{eqnarray}\label{simpleapprox}
U(x,y,z)=&-&U_0^{\rm eff}+\frac{1}{2}m\omega_r^2 \left(x^2+z^2\right)\\
&+&\mu B' \sqrt{4y^2+z_0^2}, \nonumber 
\end{eqnarray}
in which the radial confinement ($x$, $z$) is dominated by the ODT, and the axial confinement ($y$) by the QMT. The radial trap frequency is given by: 
\begin{equation}\label{radfreq}
\omega_r=\sqrt{\frac{4U_0}{mw_{0}^2}}, 
\end{equation}
where $U_0=2PC/(\pi w_0^2)$ is the ODT trap depth. The effective trap depth $U_0^{\rm eff}$ is equal to $U_0$ only for $B'= B'_{\rm lev}$, while $U_0^{\rm eff}<U_0$ for $B'< B'_{\rm lev}$ due to gravity. Expanding the trapping potential around $y=0$ gives the axial trapping frequency of the hybrid trap:
\begin{equation}\label{axialfreq}
\omega_y=\sqrt{\frac{4\mu B'}{m \left|z_0\right|}}, 
\end{equation}
which depends on $z_0$ and typically is much larger than the axial frequency of the pure ODT, $\omega_a^{\rm ODT}=\sqrt{2U_0/m y_{R}^2}$, even for a small gradient on the order of 1~G/cm. Therefore in the hybrid trap it is much easier to obtain a BEC than in a pure single beam ODT, even for a weak gradient \cite{zaiser2011smf}. 

The density distribution $n(\textbf{r})=n_0\exp\left[-U(\textbf{r})/k_B T\right]$ in the hybrid trap for temperatures much lower than the trap depth is given by:
\begin{eqnarray}\label{nhybrid}
n(x,y,z)&=& n_0 \exp\left(-\frac{\mu B'}{k_B T} \sqrt{4y^2+z_0^2}\right) \\ 
&\times& \exp\left(\frac{\mu B' z_0}{k_B T}\right)\exp\left(-\frac{m\omega_r^2 \left[x^2 + z^2\right]}{2 k_B T}\right) \nonumber, 
\end{eqnarray}
and from the condition $N=\int n(\textbf{r}) d\textbf{r}$ one finds for the peak density:
\begin{equation}\label{n0hybrid}
n_0=N\frac{\mu B' m\omega_r^2}{2\pi \left(k_B T\right)^2}F\left(\frac{\mu B' \left|z_0\right|}{k_B T}\right),
\end{equation}
where $N$ is the number of atoms and $T$ is the temperature. The function $F(x)$ is a monotonic function for which $F(0)=1$ and $F(\infty)=0$, and is discussed in Appendix \ref{app}. The overall temperature dependence is $n_0\propto T^{-2+\epsilon}$, where $0\leq\epsilon\leq 1/2$ (see Appendix \ref{app}), which lies in between that of a pure harmonic trap ($n_0 \propto T^{-3/2}$) and a pure linear trap ($n_0 \propto T^{-3}$). We also expect the value of the phase-space density $D=n_0\lambda_{\rm dB}^3$ ($\lambda_{\rm dB}=h/\sqrt{2\pi m k_B T}$ is the de Broglie wavelength) for the onset of BEC in the hybrid trap to be in between that of the pure linear trap ($1.055$) and pure harmonic trap ($1.202$) \cite{bagnato1987bec}.

In our configuration of QMT and ODT, for $B'=B'_{\rm lev}$, the trapping force in the axial direction is $2\mu B'_{\rm lev}$, compared to $\mu B'_{\rm lev}/2$ in the more common hybrid trap configuration \cite{lin2009rpo,cho2011ahp,jenkin2011bec,gotlibovych2012acs,xiong2013narbbec,xiong2013poa,kuhn2014abc,lundblad2014ool,marchant2012bec}, in which the strong axis of the QMT is vertical, while the ODT is in the horizontal plane. This provides a factor of four enhancement in the peak density in our case\footnote{Note that Eq.~\ref{n0hybrid} (and Eq.~\ref{axialfreq}) are valid if the axial confinement in the hybrid trap is provided by the strong axis of the QMT. In the geometry of the previous hybrid traps the axial confinement is provided by the weak axis of the QMT, which gives a reduction of a factor of two: i.\,e.\, $n_0=N \mu B' m\omega_r^2/[4\pi (k_B T)^2]F(\mu B' \left|z_0\right|/k_B T)$ (and $\omega_y=\sqrt{\mu B'/m \left|z_0\right|}$). In addition, for those hybrid traps the strong axis of the QMT is along the vertical axis, such that the required $B'$ for levitation is a factor of two smaller.}. Furthermore, for $^{87}$Rb, the $F=2$, $m_F=2$ state has a twice as large magnetic moment as the $F=1$, $m_F=-1$ state that gives another factor of eight in the peak density in the QMT for a given gradient. These enhancements allow for either faster evaporative cooling in the hybrid trap or efficient cooling starting with a relatively small number of atoms. 

\section{Experimental setup}\label{expsetup}

The main part of our experimental setup has already been described in Ref.~\cite{knoop2014umo}. Here we briefly summarize, and focus on the parts that have not been described earlier, i.\,e.\, MW-induced forced evaporative cooling in the QMT and the single beam optical dipole trap. In short, we load $1\times 10^9$ $^{87}$Rb atoms in a 3D-MOT from a 2D-MOT. After compression, optical molasses and optical pumping to the $F=2$, $m_F=2$ state, the atoms are loaded in the QMT at $B'$=120~G/cm. The quadrupole magnetic field for both the 3D-MOT and QMT is created by one pair of coils operating in anti-Helmholtz condition providing a gradient $B'=0.6$~(G/cm)/A.

After loading in the QMT, we apply MW-induced forced evaporative cooling, driving the $F=2$, $m_F=2$ to $F=1$, $m_F=1$ transition, resulting in an effective trap depth $U_0^{\rm eff}=(2/3)h\left(\nu_{\rm MW}-\nu_{\rm HFS}\right)\left(1-m g/\mu B'\right)$. The hyperfine splitting $\nu_{\rm HFS}$ of $^{87}$Rb is 6834.68~MHz. We generate the MW frequency $\nu_{\rm MW}$ by mixing the frequency doubled output of a tunable 80~MHz function generator with a phase locked oscillator at 6800~MHz. This circumvents the need of an expensive tunable frequency generator that reaches to at least 7~GHz. We send about 8~W MW power to a rectangular waveguide (MW horn), which is placed outside the vacuum apparatus. 

The single beam optical dipole trap at $\lambda=1557$~nm is derived from a 10~W fiber amplifier (Nufern NuAMP PSFA), seeded by a narrowband fiber laser (NP Photonics Scorpio). At this wavelength $C=1.32\times10^{-36}$ J/(W m$^{-2}$) for Rb \cite{safronova2006fdp}. For fast switching and intensity ramps we use an AOM (Crystal Technology 3165-1) operating at 165~MHz. To improve the beam pointing stability and the switching response, we drive the AOM with two frequencies, 145~MHz and 165~MHz, in which the total power of about 5~W is kept constant, and the intensity is controlled by the RF power ratio of the two RF frequencies \cite{frohlich2007tfa}. After the AOM the light is coupled into a polarization maintaining single mode fiber (OZ optics) and sent to the experimental setup. After the fiber outcoupler and a telescope the light is focused into the setup by an achromat doublet 2-inch lens with $f=400$~mm (Thorlabs, AC508-400-C). The waist $w_0$ is $39.8\pm0.3$~$\mu$m, obtained by measuring the radial trap frequency $\omega_r$ in a pure ODT (see Eq.~\ref{radfreq}). We excite the radial motion by quickly displacing the ODT beam vertically, using the piezo controlled kinematic mount (Radiant-Dyes Laser) of the last mirror before the focusing lens. This lens is on a translation stage to axially align the focus of the ODT beam with the center of the QMT, which can be done by comparing the radial trap frequency in the pure ODT and hybrid trap or minimizing axial sloshing after transfer from the hybrid trap to the pure ODT. The Rayleigh length $y_R$ is 3~mm, which is much smaller than the 4~cm distance between the glass windows of the vacuum chamber. The maximum power available at the setup is about 4~W, resulting in a maximum trap depth of 150~$\mu$K. 

A schematic of the hybrid trap configuration is given in Fig.~\ref{setup}, showing the QMT coils, ODT beam and absorption imaging beam. The axial axis of the QMT, the ODT beam and the absorption imaging beam are in the horizontal ($x-y$) plane. The ODT beam enters the setup under an angle of 11$^\circ$ with respect to the QMT axis, which leads to a reduction of the axial magnetic field gradient by a factor of $1-\sin(11^\circ)/2\approx 0.90$ in the formulas of the axial trap frequency (Eq.~\ref{axialfreq}) and densities (Eqs.~\ref{nhybrid} and ~\ref{n0hybrid}), but does not affect the vertical magnetic field gradient, relevant for levitation gradient. The absorption imaging beam crosses the ODT beam under an angle of 22$^\circ$. We use a CCD camera (Q-Imaging, Exi-Blue) with 6.45~$\mu$m pixel size and a magnification of 1. Special care was taken to compensate offset magnetic fields, such that the distance between the QMT and ODT centers does not change when ramping down the QMT gradient. We have applied RF-spectroscopy in the pure ODT to characterize and compensate offset magnetic field in all three dimensions. 

\begin{figure}
\center
\resizebox{0.45\textwidth}{!}{%
  \includegraphics{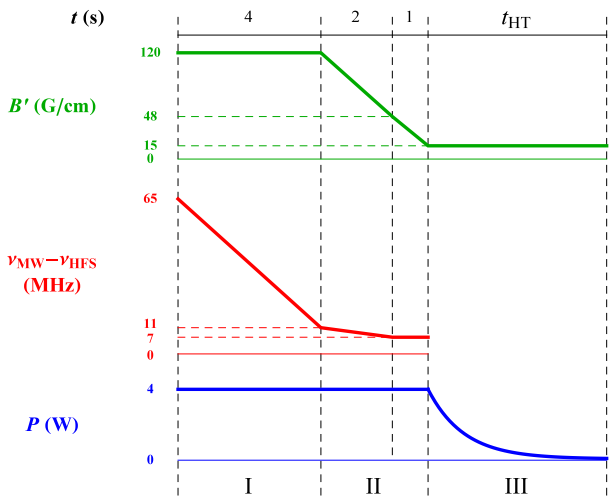}
}
\caption{(Color online) Overview of our experimental scheme for loading of and evaporation in the hybrid trap, showing the QMT gradient $B'$, microwave frequency $\nu_{\rm MW}$ and ODT power $P$. This scheme can be divided in three stages: (I) MW-forced evaporative cooling in QMT, (II) transfer from QMT to hybrid trap and (III) forced evaporative cooling in hybrid trap.}
\label{expscheme}       
\center
\end{figure}

The main experimental scheme is depicted in Fig.~\ref{expscheme}, indicating the QMT gradient $B'$, MW frequency $\nu_{\rm MW}$ and ODT power $P$. The ODT beam is on at its maximum power from the start of the QMT. In preparation of the MW-induced forced evaporative cooling, we allow for cross-dimensional thermalization for 3~s, while ramping $\nu_{\rm MW}$ from our maximum frequency of 125~MHz down to 65~MHz in 1~s and staying at 65~MHz for 2~s. Then the actual forced evaporative cooling starts by ramping down to 11~MHz in 4~s, leading to an effective trap depth of 300~$\mu$K of the pure QMT. At the end of this stage we have $2\times10^7$ atoms at 39~$\mu$K. The calculated\footnote{For a QMT the $1/e$ half width of the cloud along the weak axis is $k_B T/\mu B'$, the peak density $n_0=(N/4\pi)\left(\mu B'/k_B T\right)^3$, the mean density $\left\langle n\right\rangle=n_0/8$, and the collision rate $\gamma_{\rm col}=\sigma \left\langle n\right\rangle \left\langle v\right\rangle$, where $\left\langle v\right\rangle$ is the mean velocity and $\sigma$ the elastic cross section.} $1/e$ half width of the cloud along the weak axis is 48~$\mu$m, the peak density $1.4\times 10^{13}$~cm$^{-3}$ and the collision rate 170~s$^{-1}$.

\begin{figure}
\center
\resizebox{0.45\textwidth}{!}{%
  \includegraphics{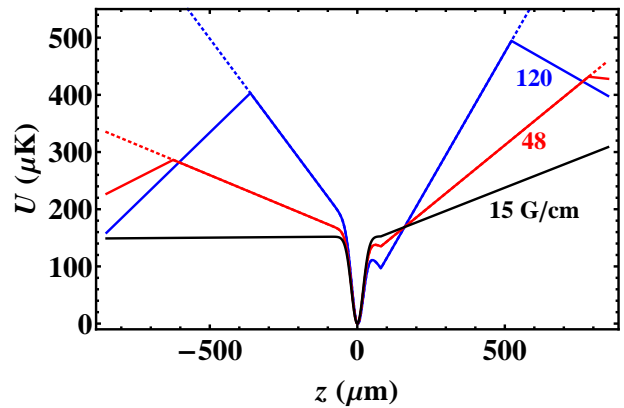}
}
\caption{(Color online) Trapping potentials of the combination of QMT and ODT, along the vertical direction, for three different magnetic field gradients $B'$, corresponding to the tight QMT (120~G/cm), decompressed QMT (48~G/cm) and hybrid trap (15~G/cm). The QMT potentials are truncated by the MW radiation, and for each particular QMT stage the situation with the lowest MW frequency $\nu_{\rm MW}$ is depicted (the dashed line represent the potentials without MW radiation). Here an offset of $z_0=80$~$\mu$m is taken, and the ODT parameters are $P=4$~W and $w_0=40$~$\mu$m.}
\label{Uloading}       
\center
\end{figure}

In the next stage we simultaneously ramp down the QMT from $B'=120$~G/cm to $B'=48$~G/cm and $\nu_{\rm MW}$ from 11~MHz to 7~MHz in 2~s, leading to an effective trap depth of 150~$\mu$K of the pure QMT. At this point we have $8\times10^6$ atoms at 20~$\mu$K. The calculated $1/e$ half width is 62~$\mu$m, the peak density $2.7\times 10^{11}$~cm$^{-3}$ and the collision rate 23~s$^{-1}$. Then $B'$ is ramped down to 15~G/cm in 1~s, just below the levitation gradient of $B'_{\rm lev}=15.4$~G/cm for $^{87}$Rb in the $F=2$, $m_F=2$ state, while $\nu_{\rm MW}$ remains at 7~MHz. Finally the MW radiation is switch off, and the ODT power is ramped down exponentially in a time $t_{\rm HT}$ for further evaporation. This scheme can be divided into three stages \cite{lin2009rpo}: (I) MW-forced evaporative cooling in the QMT, (II) transfer from QMT to hybrid trap, and (III) forced evaporative cooling in hybrid trap. The corresponding trapping potentials are depicted in Fig.~\ref{Uloading}. Alternatively, the QMT gradient is ramped down to zero in the last stage to obtain a pure ODT (see Sect.~\ref{pureODT}).

\section{Results}\label{results}

\subsection{Alignment and loading of the hybrid trap}

A crucial aspect of the hybrid trap is the radial alignment (i.\,e.\, in the $x-z$ plane) of the ODT beam with respect to the QMT center. Typically the ODT beam is placed below the QMT with $z_0 \sim w_0$. Our coarse alignment is done using in-situ absorption images to locate the positions of the hybrid trap and the QMT (without the presence of the ODT beam). For the fine alignment we scan the piezo voltages of the last mirror and measure the number of atoms loaded in the hybrid trap \cite{jenkin2011bec}. The conversion from piezo voltage to displacement is $1.9\pm 0.1$~$\mu$m/V, obtained from in-situ absorption images, such that the full scan (0-150~V) covers a range of about 300~$\mu$m, limited by the high voltage supply. 

\begin{figure}
\center
\resizebox{0.45\textwidth}{!}{%
  \includegraphics{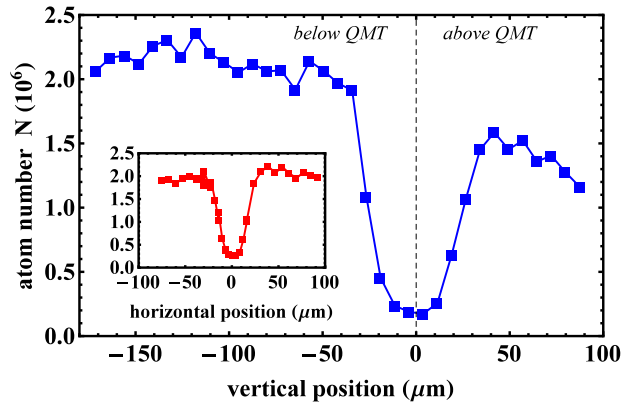}
}
\caption{(Color online) Number of atoms loaded in the hybrid trap as function of position of the ODT with respect to the QMT center in the vertical direction (main graph) and horizontal direction (inset).}
\label{piezoscan}       
\center
\end{figure}

We make use of the fact that if the ODT is located at the QMT center, the atoms will undergo Majorana spin-flips and leave the trap. A typical measurement is shown in Fig.~\ref{piezoscan}. We first do the horizontal scan (see inset) and observe a symmetric loss feature with a $1/e^2$ half width of 30~$\mu$m. We interpret the center of the loss minimum as $x=0$. At this horizontal piezo voltage, we scan in the vertical direction, again showing a clear minimum with a $1/e^2$ half width of 40~$\mu$m. Here the data clearly shows more atoms when the ODT is placed below the QMT center, and that the transfer efficiency is constant over a broad range of offsets. Still, to maximize the axial trap frequency and peak density one preferably chooses the offset as small as possible within this broad range.

\begin{figure}
\center
\resizebox{0.45\textwidth}{!}{%
  \includegraphics{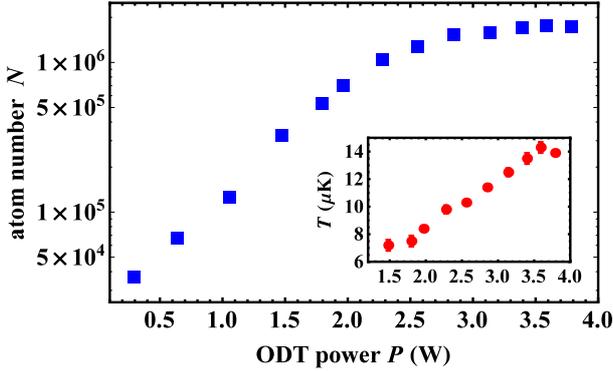}
}
\caption{(Color online) Number of atoms $N$ loaded in the hybrid trap as function of ODT power $P$. The inset shows the corresponding temperatures.}
\label{loading}      
\center
\end{figure}

After fixing the alignment at $z_0 \approx 60$~$\mu$m ($=3w_0/2$) we measure the number of atoms $N$ loaded in the hybrid trap as function of ODT power $P$. The results are shown in Fig.~\ref{loading}. For $P<2.5$~W we observe a steep increase in the number of atoms, while for $P>2.5$~W the number of loaded atoms starts to saturate. At the maximum power of 3.8~W we load $2\times10^6$ atoms at a temperature of 14~$\mu$K. The trap depth at this power is $U_0=144$~$\mu$K, such that the truncation parameter $\eta\equiv U_0/k_bT \approx 10$. The corresponding phase-space density $D=5\times10^{-3}$. For lower ODT powers (at least down to 1.5~W) we observe a linear decrease in temperature, indicating a constant $\eta$, and while the atom number decreases drastically, the phase-space density remains approximately the same. 

\subsection{Evaporative cooling in the hybrid trap}

\begin{figure}
\center
\resizebox{0.45\textwidth}{!}{%
  \includegraphics{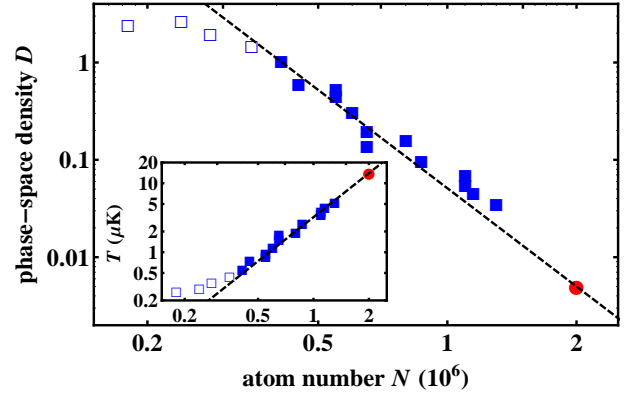}
}
\caption{(Color online) Phase-space density $D$ (main graph) and temperature $T$ (inset) as a function of the number of atoms $N$ during forced evaporative cooling in the hybrid trap (filled and oped blue squares). The dashed lines are fits based on the relationships $T\propto N^\alpha$ and $D\propto N^{-\gamma}$. The filled red circle represent the starting point of evaporative cooling. To obtain $\alpha$ and $\gamma$, the data for which $D>1$ (open blue squares) are not taken into account, for which the determination of $D$ is not correct.}
\label{evaporativecooling} 
\center
\end{figure}

After loading the hybrid trap, we ramp down the ODT power for forced evaporative cooling. In contrast to a pure ODT, the trap depth depends linearly on $P$ throughout the full range, as $B'\approx B'_{\rm lev}$. Furthermore, while the radial frequency decreases with decreasing $P$, the axial trap frequency remains constant. To investigate the evaporation efficiency we ramp down to various ODT powers and measure the number of atoms $N$ and temperature $T$, and deduce the phase-space density $D$ using Eq.~\ref{n0hybrid}. We use an approximate exponential ramp with a duration of $t_{\rm HT}=3$~s, which ensures thermalization for the full range of final values of $P$. The results are given in Fig.~\ref{evaporativecooling}, showing efficient evaporation. From the temperature as function of atom number (inset) we find $\alpha\equiv {\rm d}[\log T]/{\rm d}[\log N]=2.1(1)$, while from the phase-space density as function of atom number (main graph) we find $\gamma\equiv -{\rm d}[\log D]/{\rm d}[\log N]=3.4(1)$. To obtain these numbers we have only included the data for which $D<1$ (filled symbols). Our $\alpha$ and $\gamma$ values are similar to those observed in previous hybrid traps \cite{lin2009rpo,gotlibovych2012acs}, although our initial atom number in the hybrid trap is smaller. Here the peak density increases from $4\times 10^{13}$~cm$^{-3}$ to $8\times 10^{13}$~cm$^{-3}$, while the collision rate\footnote{The mean density in a hybrid trap is given by $\left\langle n\right\rangle=(n_0/4)F(x)/F(2x)$, where $n_0$ is given in Eq.~\ref{n0hybrid} and $x=\mu B' \left|z_0\right|/k_B T$.} decreases from 700~s$^{-1}$ to 300~s$^{-1}$.

We obtain $D\approx 1$ at a temperature of 0.5~$\mu$K and $4\times10^5$ atoms, for which we ramp down the ODT power to 150~mW. For our parameters the difference between the widths of the BEC and thermal cloud is too small to observe a bimodal distribution. The small angle of 22$^\circ$ between the ODT and imaging axes prohibits us to see the inversion of the aspect ratio as function of expansion time. The phase-space density determination for the data for which $D>1$ (open symbols in Fig.~\ref{evaporativecooling}) is not correct, because expansion of a pure thermal cloud is assumed to obtain the temperature. The lifetime of the BEC is limited by three-body losses and half of the atoms are lost in about 3~s.  

\subsection{Spin purity}

A known problem of MW-induced forced evaporation of $^{87}$Rb in the $F=2$, $m_F=2$ state, observed in harmonic magnetic traps, is the constant repopulation of atoms in the $F=2$, $m_F=1$ state, explained either by reabsorption of a MW photon while leaving the trap in the $F=1$, $m_F=1$ state \cite{haas2007ssm,xiong2010ecr,wang2011eio} or by intraspecies spin-changing collisions \cite{silber2005qdm,marzok2007uto}. This process limits the efficiency of evaporative and sympathetic cooling \cite{silber2005qdm,marzok2007uto,taglieber2008qdt}. Therefore certain cleaning procedures are applied using additional MW sweeps, which are not possible for a QMT. However, question is whether this problem occurs for a QMT at all.

\begin{figure}
\center
\resizebox{0.45\textwidth}{!}{%
  \includegraphics{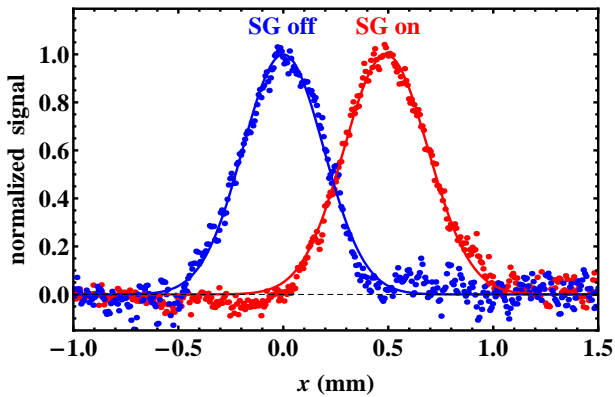}
}
\caption{(Color online) Stern-Gerlach imaging after evaporative cooling in the hybrid trap down to 0.5~$\mu$K, with 25~ms expansion time and a magnetic field gradient of 2.5~G/cm in the $x$-direction (SG on), showing an average of 10 absorption images integrated over the $z$-direction, together with the profile for which the magnetic field gradient is not applied (SG off, average of 5 images). The lines are single Gaussian distributions fitted to the data.}
\label{SternGerlach}       
\center
\end{figure}

To test the spin purity of the atoms we perform Stern-Gerlach imaging, in which we apply a magnetic field gradient in the $x$-direction after the trap is switched off, leading to a shift that depends on the magnetic moment. In Fig.~\ref{SternGerlach} we show the profiles of two absorption images, with and without an applied magnetic field gradient, both fitted by a single Gaussian distribution. The two profiles are simply shifted and have nearly the same width. Thus, all atoms are in a single Zeeman state and the Stern-Gerlach shift corresponds to the magnetic moment of the $F=2$, $m_F=2$ state. By applying a fit with two Gaussian distributions to the profile with applied magnetic field gradient, in which the second Gaussian is located at half the shift (corresponding to the magnetic moment of the $F=2$, $m_F=1$ state), we obtain an upper limit of the fraction of atoms in the $F=2$, $m_F=1$ state of 1~\%.

We are only able to do Stern-Gerlach imaging after evaporative cooling in the hybrid trap, because only then the temperature is low enough such that the Stern-Gerlach shift is at least on the same order as the width of the cloud after expansion. Therefore, we cannot experimentally test whether atoms in the $F=2$, $m_F=1$ state appear during MW evaporative cooling in the QMT. However, we would expect that atoms in $F=2$, $m_F=1$ state would be sympathetically cooled in the QMT by atoms in the $F=2$, $m_F=2$ state, transferred to the hybrid trap, and also cooled down by evaporative and sympathetic cooling in the hybrid trap. Therefore, our observation of a pure spin sample in the hybrid trap suggests that repopulation in the $F=2$, $m_F=1$ state does not occur for a QMT. 

\subsection{Pure single beam ODT}\label{pureODT}

While the hybrid trap allows for efficient transfer from the QMT and efficient evaporative cooling, for many applications a pure ODT is required. For this purpose we simply ramp down $B'$ from 48 G/cm to zero (instead of 15~G/cm), while the ODT power is maximum. In this way we end up with the same atom number as loaded in the hybrid trap ($2\times10^6$), but at a lower temperature of 10~$\mu$K (instead of 14~$\mu$K), which is mainly due to the axial decompression of the trapping potential\footnote{The lower temperature could also be explained by a reduction of the effective trap depth due to a tilt of the ODT beam in the $y-z$-plane. We have determined this tilt to be at most $0.3^\circ$, obtained by investigating the minimum power at which atoms are still trapped in the pure ODT (120~mW). The corresponding reduction of the effective trap depth for $P=4$~W is only 5\%.}.

We measure a 1/e lifetime of 105(15)~s of the trapped atoms in the pure ODT at $P=3.8$~W. This is shorter than the lifetime in the pure QMT, which we measure to be 170(10)~s and which is mainly caused by collisions with background gas (pressure in the vacuum chamber is $5\times10^{-11}$~mbar). We explain the difference between the ODT and QMT lifetimes by off-resonant photon scattering, for which at 1557~nm the rate is $6.5\times10^{-11}$$I$[Wm$^{-2}$], which is (10 s)$^{-1}$ for $P=4$~W and $w_0=40$~$\mu$m. The recoil temperature is only 0.4~$\mu$K, which is much smaller than the trap depth of 150~$\mu$K and multiple photon scattering is needed to remove an atom from the trap. Of course, for lower ODT powers the trap depth will decrease, but the scattering rate itself as well, below (100 s)$^{-1}$ for $P<400$~mW. We conclude that off-resonant photon scattering does not give a limitation to the hybrid trap and the pure ODT.

To reduce the temperature of the ultracold sample in the pure ODT one can simply reduce the ODT power. However, for a single-beam ODT forced evaporation is very inefficient because of the weak axial confinement. Therefore it is much better to do forced evaporative cooling in the hybrid trap, and ramp down the magnetic field gradient at the final ODT power. As an example, we have compared two schemes in which the final ODT power is 300~mW, at which the trap depth is 5.6~$\mu$K. For evaporation in the pure ODT, we obtain a sample of $1.2\times10^6$ atoms at 1.0~$\mu$K after 5~s, which is still not thermalized. In contrast, for evaporation in the hybrid trap, and subsequent ramping down $B'$ (linear ramp in 500~ms) we end up with a thermalized sample of $8\times10^5$ atoms at 0.5~$\mu$K, and the evaporative time can be smaller than 2~s.  In general, we obtain thermal clouds of $1\times10^6$ atoms below 1~$\mu$K.

\section{Conclusion and outlook}\label{conclusions}

We have realized BEC of $^{87}$Rb in the $F=2$, $m_F=2$ state in a hybrid trap, consisting of a QMT at 15~G/cm and a single beam ODT at 1557~nm with a waist of 40~$\mu$m and a maximum power of 4~W. In contrast to previous hybrid traps, the symmetry axis of the QMT coincides with the ODT axis, resulting in a stronger axial confinement. We find that the alignment of the ODT with respect to the QMT center is not very critical, in terms of the number of transferred atoms, as long as the vertical displacement is more than the ODT beam waist. After loading $2\times 10^6$ atoms at 14~$\mu$K from the QMT into the hybrid trap, we perform efficient forced evaporation and reach the onset of BEC at a temperature of 0.5~$\mu$K and with $4\times10^5$ atoms. We also obtain thermal clouds of $1\times10^6$ atoms below 1~$\mu$K in a pure single beam optical dipole trap, by ramping down the magnetic field gradient after evaporative cooling in the hybrid trap. We do not observe atoms in the $F=2$, $m_F=1$ state after evaporative cooling in the hybrid trap, which suggests that unwanted repopulation of this state during MW-evaporative cooling in the QMT does not take place, in contrast to harmonic magnetic traps. 

The next step is the application of the hybrid trap to an ultracold mixture of $^{87}$Rb and metastable triplet $^4$He \cite{knoop2014umo}, for which the spin purity of $^{87}$Rb is of crucial importance, as detrimental interspecies Penning ionization is expected to be only sufficiently suppressed for a spin-polarized sample \cite{byron2010sop,knoop2014umo}. An experimental appealing feature of the hybrid trap is the moderate ODT powers that is needed, in our case 3.5~W. Here the usage of an ODT at a wavelength around 1550~nm, instead of a wavelength around 1064~nm, is solely motivated by metastable triplet $^4$He, for which 1064~nm gives a blue-detuned ODT due to strong (laser cooling) transitions around 1083~nm (see e.\,g.\, Ref.~\cite{notermans2014mwf}). However, a hybrid trap using an ODT at 1064~nm would require for $^{87}$Rb an even lower ODT power of 2~W.

\begin{acknowledgement}
We gratefully thank Rob Kortekaas for technical support. This work was financially supported by the Netherlands Organisation for Scientific Research (NWO) via a VIDI grant (680-47-511) and the Dutch Foundation for Fundamental Research on Matter (FOM) via a Projectruimte grant (11PR2905). 
\end{acknowledgement}

\begin{appendix}

\section{Function $F(x)$}\label{app}

\begin{figure}
\center
\resizebox{0.45\textwidth}{!}{%
  \includegraphics{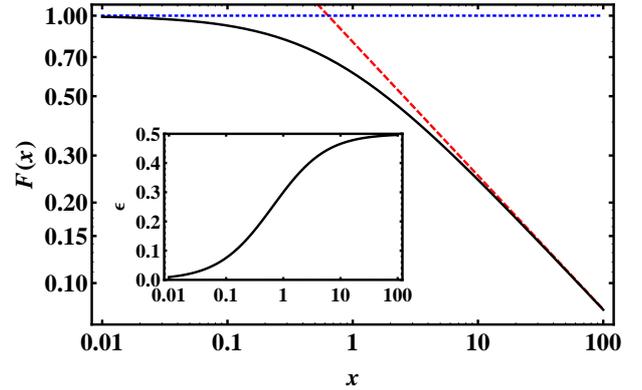}
}
\caption{The factor $F(x)$ (black solid line), relevant to calculate the peak densities (Eq.~\ref{n0hybrid}), for which $x=\mu B' \left|z_0\right|/k_B T$, together with the asymptotic solutions $F(x)=1$ (blue dotted line) and $F(x)=\sqrt{2/\pi x}$ (red dashed line). The inset shows $\epsilon$ as function of $x$, where $F(x)\propto x^{-\epsilon}$.}
\label{Ffactor}      
\center
\end{figure}

We have derived Eq.~\ref{n0hybrid} for the peak density in the hybrid trap, which contains a function $F(x)$ that is given by the integral:
\begin{equation}\label{reduction}
F(x)=\left(\int_{-\infty}^{+\infty}x\exp\left[x\left(1-\sqrt{4y^2+1}\right)\right]dy\right)^{-1}.
\end{equation}
$F(x)$ is plotted in Fig.~\ref{Ffactor}, together with its asymptotic behavior, namely $F(x)\stackrel{x\rightarrow 0}{\rightarrow}1$ and $F(x)\stackrel{x\rightarrow \infty}{\rightarrow}\sqrt{2/\pi x}$. The temperature dependence of $F(x)$, assuming $F(x)\propto T^\epsilon$, i.\,e.\, $F(x)\propto x^{-\epsilon}$, is shown in the inset of Fig.~\ref{Ffactor}, showing a smooth transition from $\epsilon=0$ to $\epsilon=1/2$.

Comparing the peak density of the hybrid trap with that of a pure harmonic trap, 
\begin{equation}
n_0^{\rm harm}=N\left(\frac{m \bar{\omega}^2}{2\pi k_B T}\right)^{3/2},
\end{equation}
where $\bar{\omega}=(\omega_r^2 \omega_y)^{1/3}$, gives:
\begin{equation}
\frac{n_0}{n_0^{\rm harm}}=\sqrt{\frac{\pi \mu B' \left|z_0\right|}{2 k_B T}}F\left(\frac{\mu B' \left|z_0\right|}{k_B T}\right)=\sqrt{\frac{\pi x}{2}}F(x)\stackrel{x\rightarrow\infty}{\rightarrow}1,
\end{equation}
which shows that only in the limit of very low temperatures and/or large offsets the peak density of the hybrid trap is equal to that of a pure harmonic trap. 

\end{appendix}

\bibliographystyle{phpf}
\bibliography{RbHybridTrap}

\end{document}